\documentclass[aps,groupeaddress,amsmath,twocolumn,amssymb,showpacs]{revtex4}
\usepackage{epsfig}

\begin{document}

\title{A few bubbles in a glass}

\author{Ludovic Berthier}
\affiliation{Laboratoire des Verres UMR 5587 CNRS,
Universit\'e Montpellier II, 34095 Montpellier, France}
  
\date{\today}

\begin{abstract}
I briefly review a recent series of papers putting forward 
a coarse-grained theoretical approach to the physics of supercooled liquids
approaching their glass transition. 
After a suitable coarse-graining, the dynamics of the liquid
is replaced by that of a mobility field, which 
can then be analytically treated.
The statistical properties of the mobility field then
determine those of the liquid. Thermodynamic, 
spatial, topographic, dynamic properties of the liquid
can then be quantitatively described within a single framework, and
derive from the existence of an underlying dynamic critical point
located at zero-temperature, where timescales and lengthscales 
diverge. 
\end{abstract}


\maketitle

\section{Why do glasses exist?}

One of the peculiarities of the field of glass transition phenomena
is that the object of study itself was already known
to mankind thousands of years ago. Therefore, 
obtaining the material to be studied is straightforward, and
Nature itself produces abundant quantities of amorphous 
solids~\cite{review1,review2,review3}.
The point is rather to understand 
{\it why} glasses exist. What is the microscopic mechanism 
that produces glassy materials? 
Does a glass state truly exist?
Can one describe the glass formation in theoretical terms
similar to the ones used for other cooperative phenomena, such as
continuous phase transitions?

Experiments performed on liquids 
supercooled through their melting transition 
towards the glass transition
have characterized static and dynamic aspects 
in much detail, so that a number of ``canonical 
features'' associated to the formation of glasses are 
well-known~\cite{review3}: broad relaxation patterns, 
extremely fast increase of relaxation times when temperature is
decreased, non-conventional hydrodynamics,
dynamic heterogeneity,
all phenomena being accompanied by no 
drastic change in the structure of the material
that simply resembles that of a normal liquid.

In a recent series of 
paper~\cite{p1,p2,p25,p3,p4,p6,p5,p8,p9,p7,p30,fss}, 
it was shown that
all these phenomena can be explained within a 
relatively simple and unique 
framework after a suitable coarse-graining mapping the density field 
to a mobility field.
After coarse-graining, the system becomes simple
enough that the statistical properties of the mobility field 
can be worked out in an essentially non-mean-field manner,
yielding analytical results and a set of quantitative predictions.
The physics is qualitatively 
explained in terms of non-trivial spatio-temporal 
correlations of the mobility field, that 
have been called bubbles. 
Analytical results suggest that these fluctuations
are critical in the sense that they reflect the 
presence of a zero-temperature dynamic critical point
where lengthscales and timescales diverge, which
was therefore proposed to be responsible for 
the experimentally observed glass transition.

The paper is organized as follows. We first explain 
how the dynamics of supercooled liquids is mapped onto 
a mobility field exhibiting non-trivial spatio-temporal correlations.
We then show that the presence of these dynamic 
bubbles is sufficient to qualitatively explain the physics.
We finally show that bubbles are the direct manifestation of dynamic critical
fluctuations that can be analyzed through the renormalization group.

\section{From molecules to bubbles}

\subsection{Coarse-graining}

\begin{figure}
\begin{center}
\psfig{file=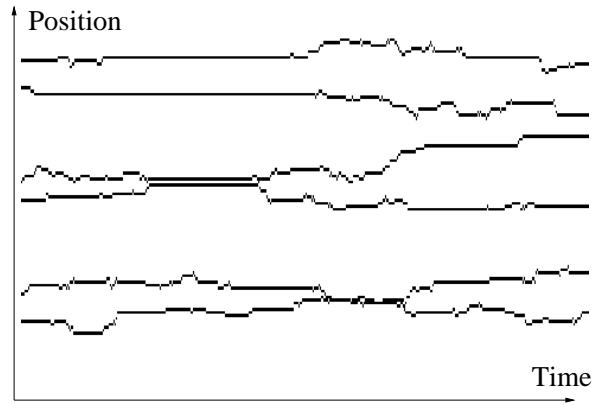,width=8.cm}
\caption{\label{prob} A piece of trajectory of six
randomly chosen probe molecules in one of the coarse-grained 
models discussed in this paper. Molecules
are caged for long periods of time (horizontal lines) 
before hopping and becoming trapped
again. Note that two molecules coming sufficiently close to one another 
have correlated dynamics. At a given time step,
some molecules are hopping while others are trapped, and
the system is therefore dynamically heterogeneous.} 
\end{center}
\end{figure}

Traditionally, simple liquids are  
described in terms of the density field, 
\begin{equation}
\rho(x,t) =  \sum_{i=1}^N \delta \big( x-x_i(t) \big),
\end{equation}
where $x_i(t)$ is the position of particle
$i \in \{1,\cdots,N\}$ at time $t$. This is because the structure 
of the liquid can be expressed in terms of 
various density correlators, whose analytical treatment 
can be found in classic textbooks~\cite{jl}.
When the temperature of a supercooled liquid is decreased towards the 
glass temperature, the static structure
barely evolves despite dramatic 
changes in the particles' dynamics, indicating
that standard liquid theory might be of modest utility here.

An amorphous solid is obtained when 
the diffusion constant of the particles 
becomes unmeasurably small.
Above the glass transition, however, 
diffusion holds at sufficiently large time and length scales.
A closer inspection of the particles' trajectories reveals
the specificities of supercooled liquids, see Fig.~\ref{prob}.
On timescales of the order of the relaxation time, $\tau_\alpha$, 
trajectories do not resemble a classic random walk. 
Rather, particles are immobile for long periods of
time, before diffusing and eventually being stuck again. 
In the traditional jargon, particles
are ``caged'', or ``trapped'', before ``hopping''.
This leads to a two-step decay 
of dynamic correlators, corresponding to a fast ``rattling'' of the
particles within their cages, and a much slower decay corresponding
to the ``structural'' decay of the liquid associated to hopping.
Explaining this two-step decay, and elucidating the properties of both
processes is the main theoretical task in the field.

\begin{figure}
\begin{center}
\psfig{file=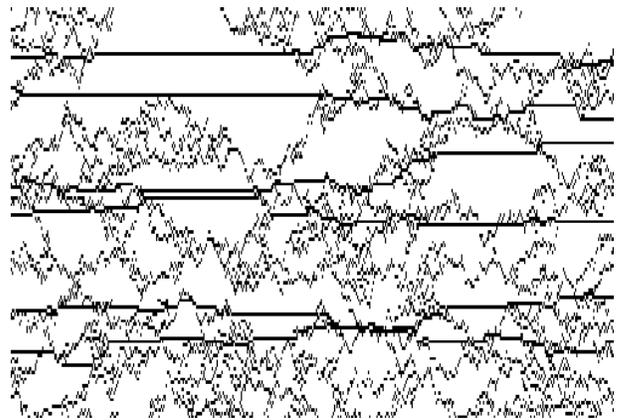,width=8.cm}
\caption{\label{prob2} From molecules
to bubbles. The molecular trajectories of Fig.~\ref{prob}
are superposed with the trajectory of an underlying binary mobility
field. Black denotes mobile regions, $\varphi=1$, white immobile
ones, $\varphi=0$.}
\end{center}
\end{figure}

The main idea of the approach described in this paper 
is to map the density field, 
$\rho(x,t)$, to a mobility
field, $\varphi(x,t)$. The mapping is illustrated in Fig.~\ref{prob2}, 
where the trajectories of Fig.~\ref{prob} are superposed 
to the dynamics of a binary mobility field, $\varphi(x,t)=0,1$.
A particle immersed in a sea of immobility (white) does not move, it
is caged. A particle can move when sitting 
on a mobile region, or ``defect'' (black). 
It turns out that the dynamics of $\varphi(x,t)$ can 
be empirically determined, leading to well-defined, solvable
models from which predictions can be made.
The obtained theory does not start from first-principles, but
it allows one to address analytically all 
desired aspects of the physics. Comparison
to experiments or simulations provides then the 
necessary check of the correctness of the approach.

Schematically, the mobility field can be obtained 
via a temporal and spatial coarse-graining~\cite{p1}, 
\begin{equation}
\varphi(x,t) = 1-\frac{1}{V_x} \int_{V_x} d^d x'
\left( \frac{
\rho(x',t) \rho(x',t+\delta t)-\langle \rho \rangle^2}{\langle \rho^2 
\rangle - \langle \rho \rangle^2} \right),
\label{cg}
\end{equation}
so that $\varphi(x,t) \approx 0$ if the local mobility is low, 
$\varphi(x,t) \approx 1$, otherwise.
The coarse-graining procedure makes use of a volume $V_x 
\sim \xi_{\rm s}^d$ centered in $x$, 
where $d$ is the space dimensionality, 
and $\xi_{\rm s}$ is the static correlation length as given
by the structure factor, and a microscopic time scale, 
$\delta t$, of the order of ps.
Doing so, we lose track of the fast vibrational 
properties of the liquid, which are thought to be irrelevant
for the glass transition problem.
In that sense, the approach is close in spirit to
inherent structure studies where real trajectories are mapped
onto local minima of the potential energy surface, also
removing fast vibrations, see Refs.~\cite{still,is1,is2}.

Note finally that more complex, e.g. vectorial, mobility
field can be defined, in order to capture a possible
local anisotropy of the dynamic facilitation~\cite{east}.
Persistence or absence of directionality of the dynamic constraint
can be shown to lead to fragile and strong behaviours, as analyzed 
in detail in Ref.~\cite{p1}.

\subsection{Link with kinetically constrained models}

After coarse-graining, Eq.~(\ref{cg}),
the properties of the mobility
field $\varphi(x,t)$ remain to be described. Here, two 
phenomenological assumptions are made.

(i) Since dynamics at low temperature is very slow, 
it is natural to expect that $\langle \varphi(x,t) \rangle \ll 1$ at low 
temperatures. Therefore, the simplest choice 
is to assume that mobile regions are so sparse that 
they do not interact. Thermodynamics is then assumed 
to be that of a non-interacting gas of point defects
of mobility.

(ii) The time evolution of $\varphi(x,t)$ makes 
use of the concept of dynamic facilitation, introduced 
long ago~\cite{fa,prl}. It is assumed that a region can 
evolve, and therefore satisfy $\partial_t \varphi \neq 0$,  
if and only if surrounded by regions of high mobility.
The simplest rule is to assume that
\begin{equation}
\frac{\partial \varphi(x,t)}{\partial t} \propto \int_{\partial V_x} 
d^d x' \, \varphi(x',t),
\label{cons}
\end{equation}
where the integral is performed over the volume $\partial V_x$
surrounding, but excluding, position $x$.
More complex rules can of course be envisaged~\cite{jackle,sollich}.

It is now obvious that this approach is close
to the kinetically constrained views put forward 
some twenty years ago in two seminal papers~\cite{fa,prl}. 
The connection is immediate if one wants to define
the simplest lattice model making use of coarse-graining
and assumptions (i) and (ii) described above. Assumption (i)
leads to 
$H = \sum_i \varphi_i$, with binary variables $\varphi_i=0,1$,
living on a regular lattice,  
and (ii) implies that $\varphi_i \leftrightarrow (1-\varphi_i)$ if and only if
at least one of the neigbours of site $i$ is mobile, $\sum_j' \varphi_j > 
0$, where the sum is over nearest neighbours.
This model is precisely one of the models defined 
by Fredrickson and Andersen (FA) in Ref.~\cite{fa}.
The pictures presented in Figs.~\ref{prob} and \ref{prob2}
are obtained in the $d=1$ version of the FA model.
The trajectories of the probe molecules are implemented as in Ref.~\cite{p2}. 

\section{Canonical features revisited}

\begin{figure}
\begin{center}
\psfig{file=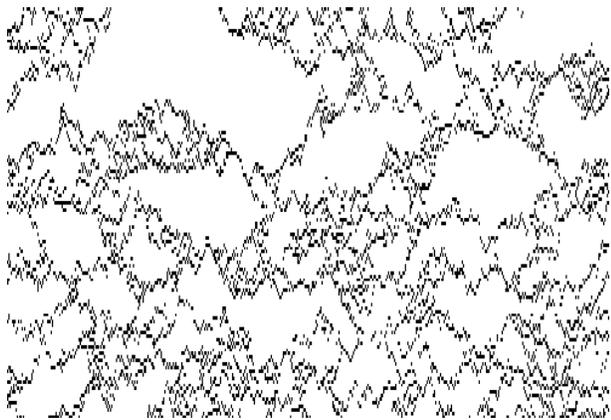,width=8.cm}
\caption{\label{mob} The mobility field of Fig.~\ref{prob2}
is shown without the molecule trajectories. We are left
with compact domains of immobility (white surfaces, the 
``bubbles'' mentioned in the title of the paper),
delimited by excitation lines shown in black.} 
\end{center}
\end{figure}

We now forget about the molecules composing the fluid, 
and consider instead the mobility field, $\varphi(x,t)$. 
In this section, we show how
the dynamical behaviour of $\varphi$ alone can lead 
to a good understanding of the physics 
of supercooled liquids and we shall 
therefore discuss Fig.~\ref{mob}.

At this stage, it is crucial to note that as far as 
qualitative physics is discussed, the $d=1$ models are just as good
as $d=3$ ones since it is known that
physics is unaltered if
dimensionality or other details like the form 
of kinetic constraint are changed~\cite{sollich}. 
Of course, these features become crucial
when quantitative comparisons to experiments or simulations
are made~\cite{p1,p25}.
This modest influence of dimensionality
applies to the coarse-grained models only, and we do not expect
$d=1$ liquids to behave as, say, the $d=1$ FA model. This is 
because even $d=1$ models are empirically defined from 
observations made in three-dimensional supercooled liquids. 
 
In Fig.~\ref{mob}, one observes diffusion, creation and coagulation
of mobility defects~\cite{p3}. Therefore, trajectories look like a mixture
of  compact  domains of immobility (white domains in Fig.~\ref{mob}),
separated by excitation lines (black lines). These white
domains are the bubbles mentioned in the title of the paper. 
We now list a number of canonical features that directly 
and generically follow
from the bubble structure of the trajectories of the mobility field.

\subsection{(Super-)Activated dynamical slowing down}

From the trajectories of the molecules shown in Fig.~\ref{prob2}, 
it it evident that the mean temporal extension of the bubbles
represents a good measure of the structural relaxation time of the system,
usually noted $\tau_\alpha(T)$, where $T$ is the temperature.
In terms of the mobility field, a natural correlator
is the local persistence, $P(x,t)$, from which 
$\tau_\alpha$ can be readily evaluated~\cite{p4}. 
It is found that $\tau_\alpha(T)$ rapidly increases
when $T$ is decreased. In the simple case of the $d=1$ FA model, 
Arrhenius behaviour is found, while super-Arrhenius
behaviour can be found in more constrained models~\cite{sollich}.
The dynamical slowing down results from the 
fact that mobility is locally needed to facilitate the dynamics, but
the mean mobility decreases at $T$ decreases. Note the 
essentially non-mean-field character of this mechanism for 
relaxation that relies on local fluctuations of the mobility
which will be hardly captured by mean-field treatments~\cite{szamel}.

\subsection{Broad distributions of relaxation times}

From Fig.~\ref{mob}, it is evident that the time extension
of the bubbles is very broadly distributed, with small
bubbles coexisting in space with very large ones.
Calculations show that the time decay
of $\langle P(x,t) \rangle$ is described, at low temperatures, by
a stretched exponential form~\cite{p3,p4}, 
\begin{equation}
\langle P(x,t) \rangle
= \exp \left[ - \left(\frac{t}{\tau_\alpha(T)} \right)^\beta \right],
\label{stretched}
\end{equation}
where the stretching exponent is constant, $\beta = 1/2$, for 
the $d=1$ FA model, but can be 
temperature dependent in more constrained dynamics. 
Therefore, the commonly used stretched exponential form
(\ref{stretched}) is a natural consequence of
the spatial heterogeneity observed in Fig.~\ref{mob}.
Moreover, detailed studies of the distributions 
of relaxation times reveal many additional features 
at intermediate times or temperatures, that
compare well to simulations and experiments~\cite{p25,p6}.

\subsection{Thermodynamics}

In this approach, thermodynamics is trivially given 
by that of a non-interacting gas of point defects. 
For the FA Hamiltonian, one gets the mean mobility
\begin{equation}
c(T) \equiv \langle \varphi \rangle = \frac{1}{1+\exp(1/T)},
\end{equation}
and the free energy, $f(T) = - T \ln{\left( 1 + e^{-1/T} \right)}$.

A traditional quantity characterizing the loss of ergodicity
taking place experimentally at the glass temperature, $T_g$, is the 
jump in specific heat $\Delta C_p$.
The configurational contribution to this jump can be readily
estimated within this approach from the free energy,
\begin{equation}
\Delta C_p(T_g) \approx \left( \frac{J}{T_g} \right)^2 {\cal N} c(T_g),
\label{cp}
\end{equation}
where $J$ sets the energy scale of the non-interacting Hamiltonian, 
and ${\cal N} \propto V_x$ is the number of molecules that contribute
to enthalpy fluctuations per mobile cell, cf Eq.~(\ref{cg}).
This shows that the concentration of mobile regions
at $c(T)$ plays again the key role, since the
jump in specific heat is given in this approach 
by the number of defects that freeze at $T_g$.
Formula (\ref{cp}) 
was successfully used in Ref.~\cite{p1} to analyze experimental
data for the jump in specific heat. Other 
thermodynamic aspects of supercooled liquids 
remain to be studied within this approach.

\subsection{Topography of the potential
energy surface}

The potential energy surface of glass-forming liquids
is an object of study which
attracted a lot of attention in recent years~\cite{review2}. 
It was shown in Ref.~\cite{p5} that the kinetically constrained
approach reviewed here is instead characterized by 
a trivial potential energy surface. However, 
this highly dimensional space is associated to a non-trivial
metric so that displacements are performed, 
at low temperatures, along its geodesics which may be highly non-trivial.
In physical terms, this implies once more that dynamical 
trajectories capture the specificities of the physics, which are
ignored if only the statics is considered. 

A detailed analysis of local minima
and saddles of the potential energy surface of kinetically constrained
models was performed in Refs.~\cite{p4,p6}, 
based on the analysis of trajectories
similar to Fig.~\ref{mob}.
The results are in nice agreement with known numerical studies
of molecular liquids. This analysis provides therefore 
a consistent physical interpretation in the ``real space'' of the 
topography found numerically in liquids.
Interestingly, this interpretation contradicts
on several key aspects the current popular views of the 
mechanisms of relaxation of supercooled liquids, see Ref.~\cite{p6}
for a more exhaustive description.

In particular, it was found that reports of several changes of mechanisms 
supposedly arising at the so-called ``mode-coupling
singularity temperature'', $T_c$~\cite{mct}, in fact result 
from a biased interpretation of numerical data, as already
discussed by several groups~\cite{review3,o1,o2,o3,o4}.
These results cast some doubts on the existence 
of a physically meaningful crossover temperature $T_c$
defined from a quantitative use of the 
mode-coupling theory of the glass transition,
and indicate  that
the topographic ``confirmations'' of a change of mechanism of 
diffusion near $T_c$ might be spurious.

Similarly, it is found that
the configurational entropy is always a
positive quantity for all $T>0$~\cite{ritort}.
Therefore there is no Kauzmann temperature, $T_K>0$, where 
the configurational entropy vanishes.
However, the shape of $S_c(T)$
explains in simple terms why numerical or 
experimental extrapolations might lead to an apparent
finite Kauzmann temperature. In this view, the ``Kauzmann
paradox'' arises simply because an invalid extrapolation
is performed.

\subsection{Spatially heterogeneous dynamics}

Dynamics in Fig.~\ref{mob} is evidently spatially heterogeneous.
The fact that kinetically constrained models are
heterogeneous was noted long ago
by Harrowell and coworkers~\cite{h1,h2,h3,h4}.

First, it is obvious that 
different parts of the same system possess, at a given time, 
very different dynamics. Secondly, it is clear that two 
sites belonging to the same bubble have similar dynamics, 
while two sites belonging to far away bubbles don't. 
This implies the existence of spatial dynamical correlations, 
and therefore of a dynamic correlation length, $\ell(T)$, 
which is given by the mean spatial extension of the bubbles. 
These correlations can be quantified 
through the following dynamic structure factor~\cite{p3,p6},
\begin{equation}
S(q,t) = \int d^d x \, e^{i q \cdot x} \left[ 
\langle P(0,t) P(x,t) \rangle - \langle P(x,t) \rangle^2 \right],
\end{equation}
which quantifies spatial correlations of the local dynamics. 
This structure factor is built from two-time, 
two-point objects, and appears to be the minimal
correlator to study bubbles, which are indeed spatio-temporal 
domains.

In the $d=1$ FA model, it is found that
$\ell(T) \sim c^{-1}(T)$, 
while more complex scaling forms can be obtained in 
other models~\cite{p25,p8,p9}.
Again, features such as ``stringy'' clusters~\cite{string}, increasing 
``four-point'' susceptibilities~\cite{silvio}, 
non-Gaussian particle displacements~\cite{weeks,blaad}, 
or results from single molecule experiments~\cite{vandenbout} can be 
easily accounted for by a detailed analysis of 
Fig.~\ref{mob}, see Refs.~\cite{p25,p3,p8,p9,p7}.

\begin{figure}
\begin{center}
\psfig{file=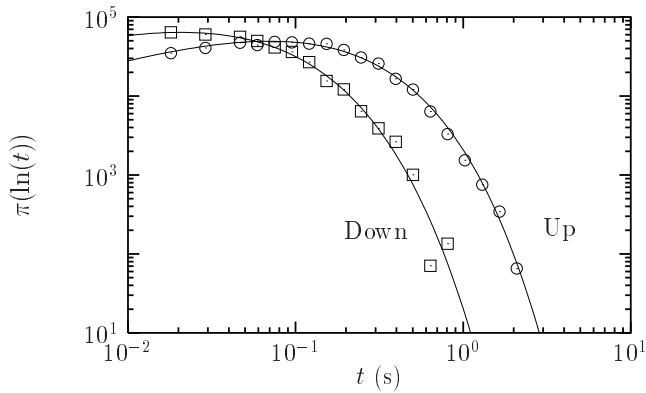,width=8.5cm}
\caption{\label{is} 
The points are distributions of flipping times for the molecular
polarizations in supercooled polyvinylacetate recorded 
at the nanometer scale with an AFM tip by Vidal-Russel
and Israeloff~\cite{isr}. The full lines are theoretical
fits using Eq.~(\ref{fit}) derived from the theoretical
approach discussed in this paper which perform much better than 
the original phenomenological fit used in~\cite{isr}.}
\end{center}
\end{figure}

As a single concrete example, we present in Fig.~\ref{is} experimental
data obtained by Vidal-Russel and Israeloff~\cite{isr}. In this experiment,
time series of molecular polarizations
are recorded with an AFM tip. This means 
that the dynamics of a very small volume is accessed, 
typically $(20\, {\rm nm})^3$~\cite{isr}. Translated in our language, 
this experimental setup probes fluctuations
at the scale of a few bubbles. The full lines in Fig.~\ref{is}
are a fit using the distribution of relaxation times
found in kinetically constrained spin models, 
\begin{equation}
\pi(t) = \frac{1}{ \tau_\alpha(T) \cal N} \left(
\frac{t}{\tau_\alpha(T)} \right)^{\beta-1} \exp \left[ - \left(
\frac{t}{\tau_\alpha(T)}
\right)^\beta \right],
\label{fit}
\end{equation}
where $\cal N$ normalizes the distribution, $\int dt \, \pi(t) =1$.
It is interesting to note that the power law prefactor was omitted
in the original experimental paper, but is necessary 
for a successfull fit of the whole experimental distributions~\cite{p4}.

\subsection{Decoupling and diffusion}

We end our short tour of the phenomenology of supercooled liquids
with the phenomenon of decoupling, which attracted a lot
of attention in the last decade. It was indeed discovered
at the beginning of the 90's that 
the temperature dependences of various dynamic measures
of the relaxation time of a given liquid do not necessarily
coincide. In particular, viscosity, $\eta \sim \tau_\alpha$, 
and diffusion constant, $D$,
have temperature dependences which result
in a breakdown of the phenomenological
Stokes-Einstein relation between these two quantities~\cite{dec1,dec2,dec3}.

Once again, Fig.~\ref{prob2} provides a simple 
explanation of this result, as was in explained in
full detail in Ref.~\cite{p2}.
We observe indeed that 
once trapped, it takes a long time to particles 
to start to move, which happens when they
are hit by a diffusing defect. However, once in movement, 
particles can make several moves very rapidly 
because they might be ``advected'' by 
the diffusing defects.
Schematically, this means that a major bottleneck for diffusion
is the first move, but once in movement diffusion
takes place on a much shorter time scale.
Since viscosity is roughly associated to the first move, 
while diffusion averages over many, 
decoupling naturally follows.

Quantitatively, it is found for the $d=1$ FA model that
$\tau_\alpha(T) \sim c^{-3}(T)$, while $D(T) \sim 
c^{2}(T)$, so that the product $D \tau_\alpha$ 
is not constant but rapidly increases when $T$ 
decreases, as found in experiments. 
Quantitative predictions for $d=3$ systems 
of various fragilities can also be found in Ref.~\cite{p2}.

This decoupling phenomenon is again a direct consequence
of the structure of trajectories shown 
in Fig.~\ref{mob}. Note that decoupling might arise
even in a case where distributions of time scales
do not broaden as $T$ decreases. In the $d=1$ FA model, for instance, 
$\pi(t,T) = \Pi(t/\tau_\alpha(T))$. This directly
contradicts the  naive argument very often put forward that 
since $D$ and $\eta$ correspond to different averages, 
respectively to $\langle t^{-1} \rangle$ and $\langle t \rangle$,
over broader and broader distributions, they have 
different $T$-dependences~\cite{dec4}. 
This is actually a crucial point since 
it may help to discriminate various approaches. 
Experimentally, decoupling is indeed found in a system
for which distributions do not broaden~\cite{dec3}.

Finally, note that naive dimensional analysis
using the Stokes-Einstein relation 
allows one to define a ``Stokes'' length scale, $a$, via~\cite{dec2,dec5} 
\begin{equation}
a \sim \frac{T}{D \eta},
\end{equation}
which decreases when $T$ decreases. This is apparently puzzling
since spatial correlations increase when $T$ decreases.
In fact, it was recently showed 
by a statistical treatment
of the trajectories shown in Fig.~\ref{prob} 
that a different dimensional analysis
in fact holds~\cite{p7}. Since $\eta \sim \tau_\alpha$,
it is possible to define a ``diffusive'' length scale, $\ell^\star$, 
as~\cite{p7} 
\begin{equation}
\ell^\star \sim \sqrt{D \eta},
\end{equation}
which increases when $T$ decreases. 
This length scale has a simple physical meaning~\cite{p7}: it is the
length scale beyond which Fickian diffusion holds. This 
new length scale was recently observed in a numerical
experiment of an atomistic supercooled liquid~\cite{ellstar}, 
and experimentally in supercooled liquid TNB~\cite{mark}. 

\section{The zero-temperature dynamic critical point}

The previous section was mainly qualitative since we 
discussed physics from a ``pictorial'' point of view, 
in order to convince readers that 
the bubble picture shown in Fig.~\ref{mob} contains
interesting physical informations when thoroughly analyzed.

In order to get a quantitative theory on top of
a pleasant physical scenario, one has to arrive 
at quantitative predictions for three-dimensional systems.
In this approach, this means being able to solve $d=3$ 
coarse-grained kinetically constrained models. Moreover, 
exact solutions of simple models have always 
played an inspiring role in statistical mechanics in general, 
the literature of glassy systems being no 
exception.

Recently, the $d=3$ version of the FA model described above 
was studied using a dynamic field-theoretic approach~\cite{p8,p9}. 
This approach is justified since we already mentioned 
several times that the physics in these models
is governed by increasing time and length scales, so that
a continuous field theory can a priori prove useful. 
Moreover, the bubbles discussed above are the result
of a reaction-diffusion type of process in terms of the point 
defects of mobility. Reaction-diffusion problems are usually described 
field-theoretically, the large scale properties 
being studied using the renormalization group~\cite{review}.
This is the path followed in Refs.~\cite{p8,p9}. 
In $d=1$, however, the problem can instead be mapped to 
interacting quantum spins, and again be studied with real space 
renormalization procedures~\cite{p30}.  

In $d=3$, the problem is solved in two steps. First, one has to derive 
a continuous field theory describing the physics. This is done
using standard techniques, starting from a discrete master equation
for the mobility, and ending with a continuous action. For 
a non-conserved mobility field, and a dynamic constraint as in 
Eq.~(\ref{cons}), one gets the following action, 
\begin{eqnarray}
\label{shift}
\lefteqn{ \mathcal{S}[\bar{\varphi},\varphi,t_0] = \int d^d x \,
\int_0^{t_0} dt {\big\lgroup} \bar{\varphi} \left( \partial_t -D_0
\nabla^2 -\kappa_0^{(m)} \right)\varphi } \nonumber \\ && +\bar{\varphi}
\varphi (\lambda_0^{(1)}+\nu_0^{(1)} \nabla^2) \varphi+\bar{\varphi} \varphi
(\lambda_0^{(2)}+\nu_0^{(2)} \nabla^2) \bar{\varphi} \varphi \nonumber \\ &&
-\bar{\varphi} \varphi (\kappa_0^{(v)} +\sigma_0 \nabla^2) \bar{\varphi}
{\big\rgroup},
\end{eqnarray}

%\begin{eqnarray}
%\label{shift}
%\lefteqn{
%\mathcal{S} [\bar{\varphi},\varphi,t_0]  =  \int d^d x 
%\int_0^{t_0} dt {\big\lgroup} \bar{\varphi} \left( \partial_t -D_0
%\nabla^2 -\kappa_0^{(m)} \right) \varphi } \nonumber \\ &&  +\bar{\varphi}
%\varphi (\lambda_0^{(1)}+\nu_0^{(1)} \nabla^2) \varphi
%\nonumber \\ &&
%+\bar{\varphi} \varphi
%(\lambda_0^{(2)}+\nu_0^{(2)} \nabla^2) \bar{\varphi} \varphi
%\nonumber \\ &&
%-\bar{\varphi} \varphi (\kappa_0^{(v)} +\sigma_0 \nabla^2) \bar{\varphi}
%%{\big\rgroup}.
%\end{eqnarray}
In this expression, $\varphi(x,t)$ can be thought of as a
continous mobility field,
while $\bar{\varphi}(x,t)$ stands for its conjugated response field.
Several coupling constants have also been introduced, see~\cite{p8}.
Since the hypothesis to derive (\ref{shift}) are the same as in the 
FA model, we expect that both models belong to the same universality class.

The large scale properties of the action are then analyzed in detail.
The action (\ref{shift}) has the form of a single species branching
and coalescing diffusion-limited reaction with
additional momentum dependent terms~\cite{review}.  
By integrating out the response
field, a Langevin equation for the evolution of
$\varphi$ is obtained. This equation has a critical point at
$c=0$, i.e. $T=0$, 
describing the crossover from an
exponential decay of mobility at finite $c$, i.e. $T>0$, to an
algebraic decay at $c=0$.  In the absence of noise,
corresponding to neglecting terms quadratic in $\bar{\varphi}$,
(\ref{shift}) admits the Gaussian exponents
$(\nu_{\perp}^G,\nu_{\parallel}^G,\beta^G)=(\frac{1}{2},1,1)$.  Here
$\nu_{\perp}$ and $\nu_{\parallel}$ control the growth of spatial
$(\xi_{\perp})$ and temporal $(\xi_{\parallel})$ length scales near
criticality, $\xi_{\perp} \sim c^{-\nu_{\perp}}$ and
$\xi_{\parallel} \sim c^{-\nu_{\parallel}}$, while $\beta$
governs the long-time scaling of the energy 
density, $n \sim c^{\beta}$.

These Gaussian power laws are modified by fluctuations which are treated
using the RG. It is found that the critical point remains at $c=0$.  There is
thus no finite temperature phase transition. 
Dimensional analysis shows that the upper critical dimension of
the model is $d_c=4$.  For $d \leq 4$ fluctuations are accounted by
studying the behaviour of the effective couplings.  
For $1 < d \leq 4$, all
interaction terms in (\ref{shift}) except for $\lambda^{(1)}$ and
$\kappa^{(v)}$ are irrelevant. It follows that our system is described
for intermediate length and time scales by the directed percolation
(DP) critical point and its associated power
laws~\cite{review}. To order $\epsilon=d_c-d$ they are $(\nu_{\perp},
\nu_{\parallel}, \beta)=(\frac{1}{2}+\frac{\epsilon}{16},
1+\frac{\epsilon}{12}, 1-\frac{\epsilon}{6})$.

This analysis implies that the slowdown is a dynamical critical slowing
down as the critical point is approached from above. Correlation time
and length scales grow as inverse powers of $c$. Thermal activation
results from $c \sim e^{-1/T}$.  Dynamical scaling is predicted to
occur when the dynamic correlation length becomes appreciably larger
than the lattice spacing. This happens therefore for temperatures
lower than $T_o$, the onset temperature for dynamic
heterogeneity~\cite{p6,Brumer-Reichman}.

These analytical results apply to systems with isotropic dynamic
facilitation. They are therefore expected to apply to strong liquids, and
to those which exhibit a crossover from fragile to strong
behaviour~\cite{p1}.  While DP
behaviour is not expected
for the case of anisotropic constraints or conserved order
parameters~\cite{p25,p9}, a zero-temperature
critical point is likely to be a generic feature of both strong and fragile
glass formers~\cite{p8,p9,Aldous-Diaconis-Toninelli-et-al}, 
so that scaling properties of liquids in their fragile
regime can also be described by a field theory~\cite{p9}.
Criticality also implies that techniques used to study 
continuous phase transitions, such as finite size scaling, 
can be employed to study the glass transition~\cite{fss}.

This field theorical approach suggests the following physical picture. The
viscosity of a supercooled liquid increases rapidly as $T$ is lowered,
because the dynamics becomes increasingly spatially correlated.  A
glass is obtained when the liquid's relaxation time exceeds the
experimental time scale.  The scaling properties of time and length
scales and therefore the physical properties of supercooled liquids
are governed by a zero temperature dynamic critical point.
It was therefore proposed that this critical point is responsible
for the experimental existence of the glass state~\cite{p8}.

\section{Conclusion}

In this paper, I have very briefly summarized a recent series of papers
dedicated to the derivation and analysis of a coarse-grained
approach to the physics of supercooled liquids. 
The main idea is to map the liquid's
degrees of freedom onto a mobility field.
After coarse-graining, simple empirical rules are used to specify
the dynamics of the mobility field, resulting
in well-defined models, that are simple
enough that a number of analytical results can be derived. 
I have shown that a wide number of physical quantities can then
be deduced from the statistical properties of the mobility
field, leading to a consistent theoretical description
of all aspects of glass transition phenomena. 

The generic scenario is that the glass formation
is driven by an underlying 
zero-temperature critical point where both 
timescales and lengthscales diverge, in the form
of non-trivial dynamic correlations, the bubbles of Fig.~\ref{mob}. 
A key temperature scale in the problem is $T_o$, 
which marks the onset of
slow dynamics. Physically, $T_o$ delimits the
the critical region, $T \in [0, T_o]$, influenced by the $T=0$ critical
point. Other popular crossover 
temperatures, such as $T_c$ and $T_K$, play no role
in this approach.
Interestingly, this approach contradicts on several important
points alternative theories, so that numerics
or experiments might clarify its status in a relatively near future.

\section*{Acknowledgments} 
This talk is based on work performed in
collaborations with David Chandler, Juanpe Garrahan and Steve
Whitelam. My work is supported by the E.U.\ Marie Curie Grant No.\
HPMF-CT-2002-01927, CNRS France, Worcester College Oxford, and 
Oxford Supercomputing Center at Oxford University.

\end{document}